# Polyurethane spray coating of aluminum wire bonds to prevent corrosion and suppress resonant oscillations

Joseph M. Izen,[a,1] Matthew Kurth[a], and Rusty Boyd[b]

[a] *University of Texas at Dallas,*
   *800 West Campbell Road, Richardson, TX 75080-3021, USA*
[b] *University of Oklahoma,*
   *440 West Brooks Street, Norman, OK 73019, USA*
   *E-mail*: `joe@utdallas.edu`

ABSTRACT: Unencapsulated aluminum wedge wire bonds are common in particle physics pixel and strip detectors. Industry-favored bulk encapsulation is eschewed due to the range of operating temperatures and radiation. Wire bond failures are a persistent source of tracking-detector failure. Unencapsulated bonds are vulnerable to condensation-induced corrosion, particularly when halides are present. Oscillations from periodic Lorentz forces are documented as another source of wire bond failure. Spray application of polyurethane coatings, performance of polyurethane-coated wire bonds after climate chamber exposure, and resonant properties of polyurethane-coated wire bonds and their resistance to periodic Lorentz forces are under study for use in a future High Luminosity Large Hadron Collider detector such as the ATLAS Inner Tracker upgrade.

---

[1] Corresponding author.

# Contents

# 1. Motivation

Aluminum wedge wire bonds are used widely in the solid-state charged-particle tracking detectors found at particle physics colliders. In commercial applications, bulk encapsulation is commonly used to protect wire bonds both from mechanical forces and corrosion. Bulk encapsulation has not been adopted at accelerators such as CERN's Large Hadron Collider (LHC) for two reasons. Plastic-based encapsulants are not qualified to withstand the ionizing radiation at a collider's collision region. Further, the coefficient of thermal expansion (CTE) mismatch between aluminum and plastic bulk encapsulants is problematic. The difference between room temperature construction and an operating temperature of -20 C is challenging, and detectors closest to the interaction point may warm up to ~50 C during beam pipe bake-out. Eschewing bulk encapsulants requires constant vigilance during assembly and integration to avoid even the slightest contact with mechanically fragile wire bonds.

    Two additional wire bond vulnerabilities have been identified, even when mechanical contact has been avoided. The CDF silicon detector group reported failures of power-carrying jumper wire bonds of their SVXII detector. Their triggered readout resulted in a periodic current through the jumpers in CDF's solenoidal magnetic field.[1] The periodic Lorentz force that resulted is understood to have been at a frequency close to the fundamental mechanical resonance of the wire bonds. Flexing fatigues the heels of bond wires, leading to bond failure. The CDF group subsequently potted wire bond feet with a bead of Sylgard® 186 Silicone Elastomer from Dow Corning. They report at least an order of magnitude improvement in the maximum tolerable current.

    The ATLAS Pixel group follow a preventative strategy similar to CDF for its disk modules and use Dymax 9001 v.3.7 to encapsulate wire bond feet.[2] A study for the ATLAS SemiConductor Tracker (SCT) explores the properties of wire bond oscillations and describes Fixed Frequency Trigger Veto (FFTV) software that is adopted by ATLAS to veto triggers at intervals that can excite a wire bond resonance. A study of ATLAS Insertable B Layer (IBL) compares wire bond resonance measurements to a finite element analysis of wire bond excitations. It concludes that the highest foreseeable currents, 100 mA/wire bond supplying

digital and analog voltage regulators of the FE-I4 front-end chip, could compromise wire bonds. Potting and coating strategies were rejected as risky, and an FFTV was implemented to avoid exciting resonances.[3] The CMS Forward Pixel Upgrade being prepared for installation in 2016 follows the CDF strategy and uses Sylgard® 186 to encapsulate the feet of wire bonds to its Token Bit Manager integrated circuit.[4]

The second wire bond vulnerability is corrosion. In the presence of liquid water, aluminum corrodes via the reaction: $Al + 3H_2O \rightarrow Al(OH)_3 + 3H_2\uparrow$. The reaction is catalyzed when halide ions are present in solution, and galvanic corrosion can contribute to corrosion when gold and aluminum are both present. Corrosion typically starts at the heel of a wire bond where the protective alumina layer on the outer surface of aluminum and the crystal structure of the aluminum are disrupted during the bonding process. In Fig. 1, two streams of hydrogen bubbles and one larger bubble are visible emanating from the heel of wire bonds that are shown immersed in deionized water. The printed circuit traces were plated with the ENIG process. A forensic signature of a wire-bound corrosion attack on a dry board is a white aluminum hydroxide residue such as those shown in Fig. 2. Aluminum hydroxide residues are best imaged with oblique lighting. The staff of CERN's QART Lab carried out a survey of printed circuit boards from an assortment of CERN experiments and labs. They were able to provoke corrosion with deionized water in six of eight boards that were tested.[5]

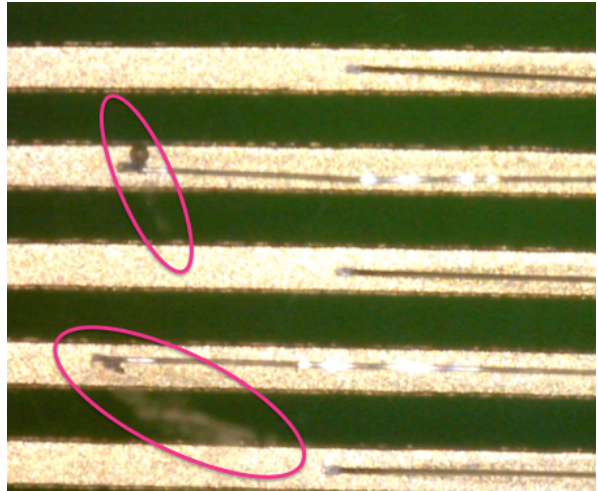

Fig. 1. Aluminum wire bonds in deionized water. Hydrogen plumes and a larger bubble emanate from wire bond heels.

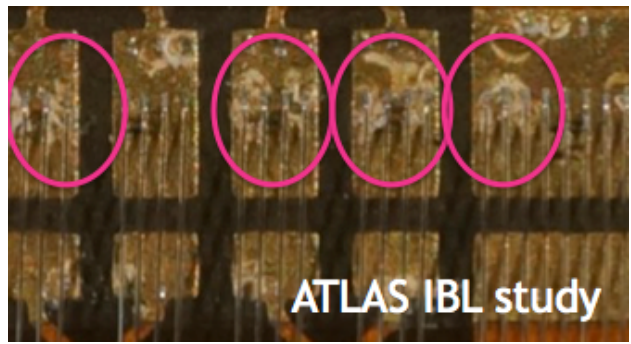

Fig. 2. White residues produced by corrosion .

Ordinarily, water should not be present on the wire bonds of a silicon tracking detector, and the detectors operate in controlled, humidity-free environments. However, the operating temperature of pixel and silicon strip modules is cold enough to risk condensation or even frost if humidity is accidently introduced. Thermal stressing during quality assurance testing causes the same potential risk.

## 2. Polyurethane coatings

Thin polyurethane coatings of wire bonds applied with an atomized spray are under investigation as a novel approach to encapsulation. The entire length of a wire bond receives a coating, not just the wire bond foot. Polyurethanes are copolymers formed by reacting an isocyanate with a polyol. Long, flexible polyol segments and low crosslinking makes polyurethane elastic, whereas shorter chains and crosslinking between chains yields a harder, more durable polymer.

The efficacy of polyurethane coatings is being tested for operating conditions anticipated for the ATLAS Inner Tracker (ITK) upgrade's pixel and strip trackers [6] being developed for the High Luminosity-Large Hadron Collider. The polyurethane selected for this study, Cellpack D 9201, has been used previously at CERN to coat printed circuits.[7] It is marketed as an electrical insulator.

## 2.1 Spraying polyurethane

Cellpack D 9201 polyurethane is sold in spray cans and as a liquid. Previous attempts to apply a coating with a spray can yielded non-uniform coatings [3] with droplets coalescing on wire bonds such as shown in Fig. 3. Satisfactory coatings were achieved with Cellpack liquid applied using a Paasche Talon Dual Action Airbrush TG2L and a 0.65mm needle and nozzle designed for spraying more viscous paints. Control of droplet size with air pressure and flow rate is crucial. When droplets are too fine, they become tacky en route to the wire (Fig. 4a). If the spray is too heavy, droplets coalesce on wires, just as if they had been applied with a spray can (Fig. 4c). In the Goldilocks zone, uniform coatings are achieved (Fig. 4b).

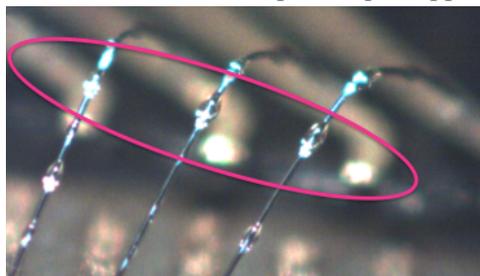

Fig. 3. Polyurethane coatings from a spray can.

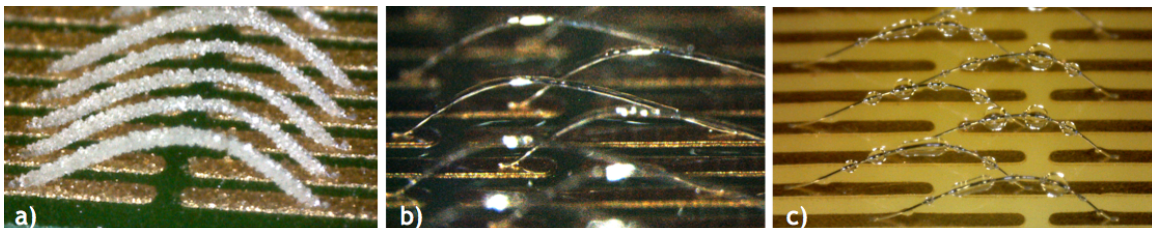

Fig. 4. Examples of polyurethane spraying: a) Spray too fine. Droplets dry before reaching wire. b) Hermetic coatings result when spray is just right. c) Spray too heavy. Droplets coalesce.

The polyurethane coating thickness is built up by passing the spray over the sample many (tens) of times with just a few seconds between spray passes. Coatings with an outer diameter from 35µ to 100µ on a 25µ diameter wire have been obtained. Wires bonds are hermetically coated, but the coating is asymmetrically deposited on the side of the wire facing the spray (Fig. 5). The wires were sprayed from each end of the bond to ensure coverage of both "ankle" regions above the wire bond feet. The polyurethane that is deposited over the feet and heels of a wire bond blends into the layer that covers the circuit board.

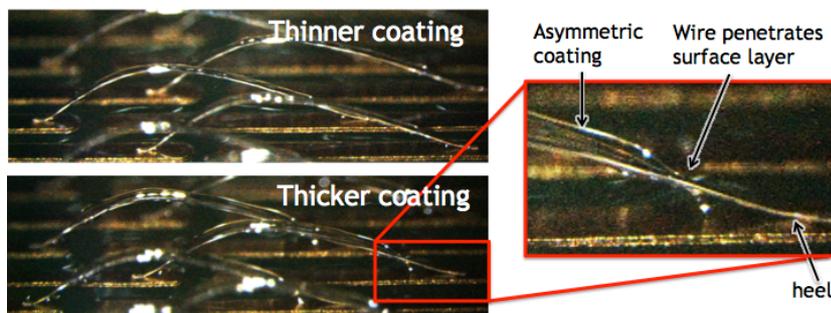

Fig. 5. Multiple spray passes build up the desired coating thickness.

**2.2 Corrosion resistance**

Ten wire bond test boards, each having seventeen 25μ diameter wire bonds, were coated with polyurethane to outer diameters of 35μ to 100μ. No hydrogen bubbling or corrosion residues were present after 20 minutes in deionized water. The samples were cycled 350 times in a climate chamber between -30 C and +50 C over a 4-day period. Testing in deionized water still failed to provoke corrosion, and no bonds failed due to the thermal cycling.

**2.3 Resistance to resonant Lorentz forces**

Resonant oscillation testing of wire bonds is carried out within the apparatus depicted in Fig. 6. A water-cooled magnet produces a horizontal magnetic field of up to 1.7 T to simulate ATLAS' 2T solenoidal field. A test board with wire bonds is oriented with its normal parallel to the field to simulate the ITK disk/endcap geometry. A waveform generator sends a 50% duty cycle square-wave current of up to 180 mA p-p through one wire bond. The Lorentz force on a 180 mA current in a 1.7 T field is 50% greater than the worst-case force anticipated on an ITK FE-I4 voltage regulator supply wire bond. The transverse periodic Lorentz force drives a guitar-like oscillation. Wire bonds are imaged with a microscope camera, and the details of wire bond motion are visualized with an LED strobe.

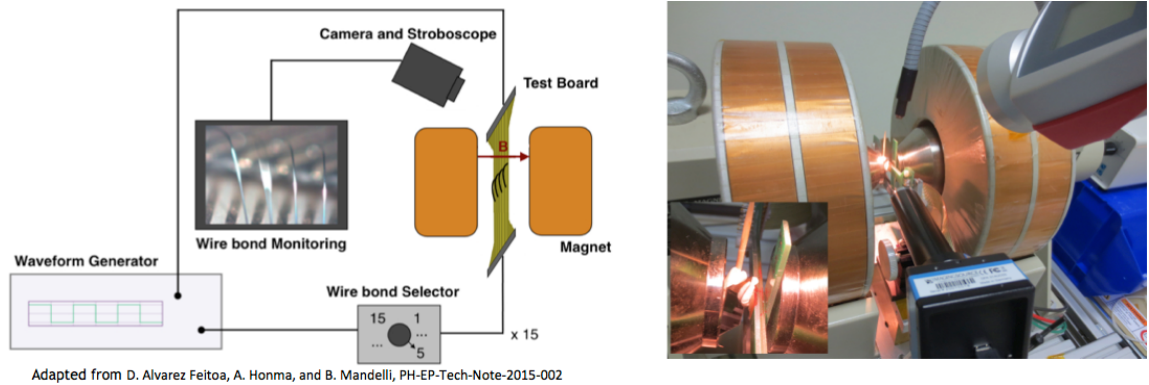

Fig. 6. Test set-up for measuring wire bond resonance properties.

Video recordings reveal that the surface layer of polyurethane surrounding and covering the wire bond foot and ankle regions keeps the most vulnerable regions immobilized. Flexing occurs higher along the wire shaft where the coated wire emerges from the surface layer. The immobilization of the wire bond foot in Fig. 7 is demonstrated by the addition and subtraction of video frames. In subtracted frames, locations that are not moving appear as a neutral gray, whereas displaced images are dark or light. Polyurethane coatings achieve the same foot immobilization as potted wire bond feet, but resonance studies suggest an additional protection mechanism.

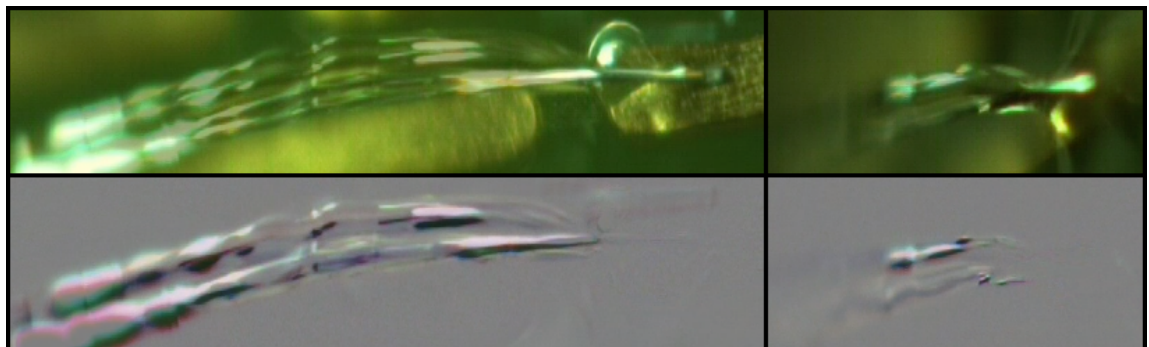

Fig. 7. Video frames have been added (top) and subtracted (bottom) to show regions of movement (light and dark) and the fixed regions (gray) that include the wire bond heel.

The low-amplitude resonance frequency was determined for each 2.8 mm wire bond, and its Q-value is inferred from the width of its resonance excitation curve. Coatings are applied manually with an airbrush, so some variation in coating thickness and resonance frequency is seen, even amongst wires on the same test board. The variation is greater on more heavily coated wire bonds. A representative sample of measurements is summarized in Table 1. Uncoated wire bonds resonate at ~11.8 kHz and typically have a Q-value of 92. A wire bond test sample having beads of polyurethane encapsulating just the wire bond feet have resonance frequencies near 15 kHz due to the shortening of the flexing region. The bond wire can be understood to be a cantilever oscillator with a frequency that increases as the flexing portion is shortened. The polyurethane-coated samples behave differently. As the coating thickness increases, the resonance frequency of wire bonds decreases due to the increase in mass along the cantilever, and the Q-value drops due to the conversion of mechanical to heat in the flexible polyurethane coating. Resonance frequencies as low as 8 kHz are seen, but further thickening of the coating leads to an increase in resonance frequency as the stiffness of the coating becomes increasingly important. The lowest Q-values are measured for wires with coated diameters around 100μ and resonance frequencies exceeding that of uncoated wire bonds. The Q-values of coated wires are an order of magnitude reduced from their uncoated counterparts, and provide significantly better damping than potted wire bonds.

A small number of wire bonds were tested to destruction. As the current is increased in increments, and resonance frequency trends gently downward as expected. The generator frequency was adjusted at each current step to the resonance frequency. A dramatic decrease in resonance frequency signals the approach of the elastic limit of a wire bond. Once the elastic limit is reached, a bond sometimes breaks during the search for the new resonance frequency, so only a current range can be quoted in Table 1. The ability to withstand larger peak-to-peak currents is anticorrelated with Q-value. Uncoated wire bonds break easily with a peak-to-peak current of a few mA at their resonance frequency. At the other end of the spectrum, four heavily-coated wires were bullet-proof; they withstood a square wave at their resonance frequencies corresponding to a peak-to-peak current 50% greater than the maximum expected for the ITK for at least an hour. The bond wire in Fig. 7 withstood that current for 38.5 hours!

| Sample | $f_{res}$ [kHz] mean (range) | Q mean (range) | $I_{p-p}$ [mA] to destroy |
|---|---|---|---|
| uncoated, $N_{wires}$=17 | 11.78 (11.68 – 11.97) | 92 (69 – 117) | 4 one wire tested |
| potted, $N_{wires}$ = 8 | 14.95 (13.80 – 16.17) | 68 (60 – 77) | 12 – 15 one wire tested |
| Lighter coating, $N_{wires}$ = 15 | 9.28 (8.88 – 9.76) | 36 (26 – 46) | 32 – 40 one wire tested |
| Heavier coating, $N_{wires}$ = 8 | (8.1 – 14.1) | (7 – 14) | $f_{res}$ = 13.3 kHz wire survives 38.5 hours @ 180 mA p-p, 1.7 T |

Table 1. Resonance properties of uncoated 25μ, 2.8 mm long, 25μ diameter bond wires that are uncoated, potted with polyurethane, and having lighter or heavier polyurethane coatings.

**2.4 Irradiation test**

A preliminary irradiation study was carried out on a wire bond sample having an ~100μ outer diameter polyurethane coating. The sample was irradiated with 27 MeV protons corresponding to 10.3 MGy. For comparison, the maximum ionizing radiation dose for the inner edge of the first ITK pixel endcap is expected to be 0.9 MGy, [6] so the dose is an order of magnitude larger than is anticipated. The resonance frequency and Q-value before and after irradiation are presented in Table 2. The increase in resonance frequency and Q-value are consistent with radiation hardening of the polyurethane. Nevertheless, all wires in this sample have been shown to withstand a square wave at resonance corresponding to a peak-to-peak current 35% greater than the maximum expected for the ITK.

| 100μ coating | Before Irradiation | | After Irradiation | | |
|---|---|---|---|---|---|
| Wire Number | $f_{res}$ [Khz] ±0.1 | $Q_{before}$ ±7% | $f_{res}$ [Khz] ±0.1 | $Q_{after}$ ±7% | $Q_{before}/Q_{after}$ ±10% |
| 2 | 9.8 | 15.0 | 11.3 | 35.9 | 2.4 |
| 3 | 9.8 | 12.5 | 11.6 | 32.1 | 2.6 |
| 4 | 9.9 | 15.1 | 11.9 | 33.0 | 2.2 |
| 10 | 12.8 | 12.2 | 16.2 | 32.5 | 2.7 |
| 12 | 13.5 | 14.7 | 17.1 | 31.7 | 2.2 |
| 13 | 13.5 | 12.1 | 16.9 | 34.9 | 2.9 |
| 15 | 12.5 | 13.6 | 16.7 | 40.3 | 3.0 |
| 16 | 11.4 | 14.5 | 13.7 | 30.3 | 2.1 |

Table 2. Resonance properties of polyurethane-coated wire bonds before and after irradiation.

**3. Conclusions and future work**

We have demonstrated that an atomized polyurethane spray can deliver uniform coatings to protect wire bonds from corrosion and resonant Lorentz forces. Polyurethane performs well in corrosion and damping studies. Initial irradiation tests are encouraging, but they need to be repeated at the lower maximum doses anticipated for the ITK pixel and ITK strip endcaps. We aim to repeat resonance measurements at -20 C to confirm oscillation damping is sufficient at the actual operational temperature of the ITK.

**Acknowledgments**

The authors wish to thank the staff of CERN's Departmental Silicon Facility: A. Honma, F. Manolescu, and I. McGill, for the use of QART Lab facilities, for bonding samples, and for many useful suggestions. Thanks to B. Mandelli for paving the way with IBL wire bond resonance measurements. Thanks to J. Wilson and P. Dervan for irradiation of samples at the University of Birmingham Cyclotron. This work was supported by US Dept. of Energy grants DE-SC0010384 and DE-SC0009956.